\documentclass[5p,times]{elsarticle}
\usepackage{algorithm}
\usepackage{array}
\usepackage{textcomp}
\usepackage{stfloats}
\usepackage{url}
\usepackage{verbatim}
\usepackage{graphicx}
\usepackage{multirow}
\usepackage{booktabs}
\usepackage{soul}
\usepackage[acronym]{glossaries}
\usepackage{xspace}
\usepackage{subcaption}
\usepackage{tikz}
\usetikzlibrary{arrows, shapes, positioning}
\usetikzlibrary{arrows.meta, shapes.geometric, positioning, svg.path}
\usepackage{algpseudocode}
\usepackage[most]{tcolorbox}
\usepackage{svg}
\usepackage[hidelinks]{hyperref}
\usepackage[capitalize]{cleveref}
\usepackage{wrapfig}

\definecolor{myblue}{HTML}{4C72B0}   
\definecolor{mygreen}{HTML}{55A868}  
\definecolor{myred}{HTML}{C44E52}    
\definecolor{mypurple}{HTML}{8172B2} 
\definecolor{myyellow}{HTML}{F0E442}
\definecolor{mygold}{HTML}{FFD700}
\definecolor{mysoftgold}{HTML}{E5C07B}

\makeatletter
\AddToHook{cmd/appendices/before}{\def\cref@section@alias{appendix}\def\cref@subsection@alias{appendix}}
\makeatother

\usepackage{ifthen}
\usepackage{amssymb}

\newboolean{showcomments}
\setboolean{showcomments}{true}

\ifthenelse{\boolean{showcomments}}
{
  \newcommand{\nb}[2]{
    \fbox{\bfseries\sffamily\tiny#1}
      {\sf\footnotesize\textit{#2}}
  }
}
{
  \newcommand{\nb}[2]{}
}

 \newacronym{jagt}{JAGT}{Judgment-Aggregated Ground Truth}
\newcommand{\jagt}{\gls{jagt}\xspace}
\newacronym{ssd}{SSD}{Steady-State Detection}
\newcommand{\ssd}{\gls{ssd}\xspace}
\newcommand{\ssdns}{\gls{ssd}}
\newacronym{kssd}{KSSD}{Kelly's Steady State Detection}
\newcommand{\kssd}{\gls{kssd}\xspace}
\newacronym{kbkssd}{KB-KSSD}{Kernel-based Kelly's Steady State Detection}
\newcommand{\kbkssd}{\gls{kbkssd}\xspace}
\newcommand{\kbkssdns}{\gls{kbkssd}\xspace}
\newacronym{cpssd}{CP-SSD}{Change-Point Steady State Detection}
\newcommand{\cpssd}{\gls{cpssd}\xspace}
\newcommand{\ie}{\emph{i.e.,}\xspace}

\newcommand{\SoBigDataITAck}{European Union -- NextGenerationEU -- National Recovery and Resilience Plan (Piano Nazionale di Ripresa e Resilienza, PNRR) -- Project: ``SoBigData.it -- Strengthening the Italian RI for Social Mining and Big Data Analytics'' -- cod. IR0000013 -- D.D. n. 3264 del 28/12/2021\xspace}
\newcommand{\rechargeAck}{Italian Government (Ministero dell'Università e della Ricerca, PRIN 2022 PNRR) -- cod. P2022SELA7: ''RECHARGE: monitoRing, tEsting, and CHaracterization of performAnce Regressions`` -- D.D. n. 1205 del 28/7/2023\xspace}
\newcommand{\icscAck}{European Union -- NextGenerationEU -- National Recovery and Resilience Plan (Piano Nazionale di Ripresa e Resilienza, PNRR) -- Project: ``ICSC – Centro Nazionale di Ricerca in High Performance Computing, Big Data and Quantum Computing'' -- cod. CN00000013 -- D.D. n. 3138 del 16/12/2021\xspace}
\newcommand{\mattersAck}{European Union -- NextGenerationEU -- National Recovery and Resilience Plan (Piano Nazionale di Ripresa e Resilienza, PNRR) -- Project: “Modeling and Analyzing Tradeoffs beTween cybERSecurity and other quality attributes in Cyber-Physical Systems” (MATTERS), in the context of Spoke 8 – Risk Management and Governance (SERICS) – “SEcurity and RIghts In the CyberSpace” (CUP: J33C22002810001)\xspace}

\newtcolorbox{rqbox}{
  enhanced,
  boxrule=0pt,frame hidden,
  borderline west={4pt}{0pt}{black!50},
  colback=gray!5,
  sharp corners
}
 
\journal{Future Generation Computer Systems}

\makeatletter
\renewcommand{\subsubsection}{\@startsection{subsubsection}{3}{0pt}{0.5em}  {0.25em}    {\normalfont\normalsize\itshape}} 

\renewcommand{\thesubsubsection}{\alph{subsubsection})} 

\renewcommand{\@seccntformat}[1]{\ifx#1subsubsection\relax
    \thesubsubsection\quad \else
    \csname the#1\endcsname\quad \fi
}
\makeatother

\begin{document}

\begin{frontmatter}

\title{A Kernel-Based Approach for Accurate Steady-State Detection in Performance Time Series}

\author[first]{Martin Beseda\corref{cor1}}\author[first]{Vittorio Cortellessa}\author[first]{Daniele {Di Pompeo}}\author[first]{Luca Traini}\author[first]{Michele Tucci}

\affiliation[first]{organization={University of L'Aquila},
            addressline={via Vetoio (Coppito), 1 -L'Aquila}, 
            city={Earth},
            postcode={67100},
            country={Italy}}

\cortext[cor1]{Corresponding author. E-mail address: martin.beseda@univaq.it}

\begin{abstract}
This paper addresses the challenge of accurately detecting the transition from the warmup phase to the steady state in performance metric time series, which is a critical step for effective benchmarking. The goal is to introduce a method that avoids premature or delayed detection, which can lead to inaccurate or inefficient performance analysis. The proposed approach adapts techniques from the chemical reactors domain, detecting steady states online through the combination of kernel-based step detection and statistical methods. By using a window-based approach, it provides detailed information and improves the accuracy of identifying phase transitions, even in noisy or irregular time series. Results show that the new approach reduces total error by 14.5\% compared to the state-of-the-art method. It offers more reliable detection of the steady-state onset, delivering greater precision for benchmarking tasks. For users, the new approach enhances the accuracy and stability of performance benchmarking, efficiently handling diverse time series data. Its robustness and adaptability make it a valuable tool for real-world performance evaluation, ensuring consistent and reproducible results.
\end{abstract}
 
\begin{keyword}
steady state \sep convolution \sep kernel \sep time series \sep warm up \sep performance
\end{keyword}

\end{frontmatter}

\section{Introduction}
\label{sec:intro}
Time series of performance metrics collected on running systems are notoriously subjected to a two-phase behavior: (i) in the first phase, namely \emph{warm-up}, the metric trend is irregular, mainly due to instability of resource assignment and/or runtime optimizations on the executed code; (ii) in a second phase, namely \emph{steady state}, the collected values of metrics may achieve a certain degree of regularity over time. For obvious reasons, the warm-up section of the time series is excluded from the analysis of collected data. 

A major (still unsolved) problem in performance benchmarking/testing is accurately identifying the phase change in a time series, i.e. the transition from warm-up to steady state. The complexity of this task is due to several factors, like transient stable behaviors that can be erroneously interpreted as achieved steady states, or odd distributions of outliers across the whole time series. However, this task remains crucial for the effectiveness of performance benchmarking/testing. Indeed, a too-early identification of the phase change leads to the risk of accounting for unstable values in the analysis process, whereas a too-late identification leads to missing accurate values and/or spending useless additional testing time to achieve an adequate amount of stable measures for the analysis process.
Despite its complexity, several approaches have been introduced in the last few years for addressing this problem, with different degrees of accuracy and effectiveness, ranging from statistical approaches to machine learning methods. Techniques like changepoint detection algorithms \cite{truong2020,ieee2022} or hybrid models combining statistical tests \cite{liu2017,turan2023,liu2019} have seen increasing use in benchmarking scenarios. However, accurately capturing the phase transition remains an open issue, particularly when faced with irregular time series or noise-heavy environments, as is common in system performance metrics.

In this paper, we consider a modification of approaches that are traditionally adopted for \ssd in time series from the chemical reactor domain \cite{kelly2013steady}. As such, the presented method \kbkssd is able to monitor steadiness in an ``online'' manner and to provide some information on the ``window'' basis, i.e. the user can obtain more detailed information about the character of the analyzed time series.

We tailor the original approaches to the case of performance metric time series, thus introducing a novel technique that we apply to a dataset of 586 time series. We compare the results obtained with this technique to the ones obtained by \cpssd\,\cite{TrainiCPT23}, considering it a state-of-the-art technique with the same major use cases. 

For the sake of comparison, we have built a ground truth on the considered dataset using judgment aggregation. With that approach, 407 series were considered steady and further evaluated for the steady-state starting indices. This approach is described in detail later on.

In this paper, we aim to address the following research questions.

\textbf{RQ1. How does \bf{the new technique} compare to existing ones in terms of the accuracy of steady state detection?}
    
This is a comparison against the benchmark of the old and new methods from two points of view: a) how good the methods are at classifying time series as steady or non-steady; b) how distant (in terms of iterations and/or time) the two methods are from the ground truth.

\textbf{RQ2. How does \bf{the new technique} effectiveness vary while modifying its main parameters?}
The analysis of the model's sensitivity w.r.t. the change in its parameters. The considered parameters are the size of the sliding window for ``smoothing'' the outliers, the size of the sliding window checking steadiness in \kbkssd, the t-test critical value used in \kbkssd, the size of the sliding window for kernel-based step detection and the probability threshold for \kbkssd.

The presented method demonstrates superior performance compared to \cpssd in several key aspects. When evaluated against a ground truth set of 586 time series, \kbkssd achieved a slightly higher number of agreements, showing a lower rate of false negatives (0.86\%) compared to \cpssd (4.6\%). Further analysis of \ssd accuracy revealed that \kbkssd's absolute error was 14.5\% lower than that of \cpssd. Additionally, \kbkssd displayed more stable behavior, with a significantly lower standard deviation of signed errors.

The paper is organized into five sections following the introduction. Section \ref{sec:background} defines key concepts such as \textit{microbenchmarking}, \textit{warm-up phase}, and \textit{steady state}, which lay the groundwork for the subsequent analysis. Section \ref{sec:approach} presents a detailed explanation of a novel kernel-based approach for detecting the steady state, offering insights into its methodology and underlying principles. The discussion of experimental results takes place in Section \ref{sec:results}, where we not only present our findings but also explain the process of constructing the ground truth set used to validate our approach. Section \ref{sec:threats} addresses potential threats to the validity of our results, suggesting possible strategies for overcoming or mitigating these concerns in future work. Finally, the paper concludes with Section \ref{sec:conclusion}, where a summary of the key results and insights, highlights where \kbkssd outperforms \cpssd method and offers directions for further research.

The code can be accessed on the public GitHub repository\footnote{\url{https://github.com/MartinBeseda/steady-state.git}}, with the relevant scripts released as v1.0.0\footnote{\url{https://github.com/MartinBeseda/steady-state/releases/tag/v1.0.0}}. This release was also published on Zenodo\,\cite{beseda_2025_15005959}.
 \section{Background and Related Work}
\label{sec:background}

\subsection{Java microbenchmarking}

The Java Microbenchmark Harness (JMH) is the standard framework for writing and executing microbenchmarks for Java software. It helps developers create and run microbenchmarks that measure the performance of Java code segments, such as methods. 

To tune the reliability of measurement, JMH supports three levels of repetitions: forks, iterations, and invocations. Invocations, the most fine-grained level of repetition, involve running microbenchmark executions continuously within a predefined time period, referred to as an iteration. A fork encompasses a series of iterations and is executed on a completely fresh Java Virtual Machine (JVM). 

Within a fork, there are two types of iterations: warmup and measurement. A warmup iteration is designed to prepare the JVM for actual execution, and it ensures that measurements are not taken when the JVM is in a cold state (i.e. when it is still performing internal processes, such as class loading). In contrast, a measurement iteration is where JMH collects performance metrics, such as average execution time, for the microbenchmark.

To mitigate the effects of confounding factors, iterations should be repeated multiple times on fresh JVMs~\cite{Georges2007, Kalibera2013, Barrett2017,TrainiCPT23}.

\subsection{Warmup and steady state}
In the first phase of their execution, Java microbenchmarks are slowly executed by the JVM. Over time, the JVM identifies frequently executed loops or methods and dynamically compiles them into optimized machine code. This process typically speeds up subsequent microbenchmark executions. Once a dynamic compilation concludes, the JVM is considered warmed up, and the benchmark is said to execute in a \emph{steady state of performance}.

The primary goal of Java microbenchmarking is to evaluate the steady-state performance of Java applications. A common method to achieve this involves running the benchmark multiple times and discarding the first executions, which correspond to the warmup phase, to avoid skewing results. However, using a fixed number of executions does not guarantee that the warmup phase has fully ended. To address this, researchers have developed data-driven approaches to rigorously determine the end of the warmup phase. Notable methodologies include those by Georges et al. \cite{Georges2007} and Kalibera and Jones \cite{Kalibera2013}. Georges et al.'s approach relies on preset thresholds for the coefficient of variation to detect the end of warmup, while Kalibera and Jones use data visualization techniques (e.g., autocorrelation function plots, lag plots, and run-sequence plots). Despite their utility, both methods have drawbacks: Georges et al.’s heuristic often fails to reliably identify the end of warmup~\cite{Kalibera2013}, while Kalibera and Jones’ visualization-based approach is primarily manual, leading to significant limitations such as: (i) susceptibility to human error or disagreement, and (ii) lack of automation, which hampers scalability.

To address these challenges, Barrett et al. \cite{Barrett2017} introduced an automated method leveraging change point detection~\cite{Eckley2011}. This technique offers a more rigorous alternative to Georges et al.’s heuristic while enabling full automation, unlike Kalibera and Jones’ manual approach. The method employs a standard change point detection algorithm, PELT~\cite{Killick2012}, to identify changes in the execution time of the benchmark. These detected changes are post-processed to determine if and when a benchmark reaches steady-state performance. To date, this approach represents the state-of-the-art for \ssd.
 \section{Steady state detection using a probabilistic approach}
\label{sec:approach}
In this section, we describe the way time series are pre-processed, together with the combination of methods used to detect (un)steadiness, and the way we evaluate the model sensitivity w.r.t. its parameters' changes. 

\subsection{Smoothing Approach}\label{sec:smoothing}
Time series data are often noisy and contain significant outliers, which can obscure meaningful trends. To focus on the underlying trends and prevent misinterpretation, especially in cases where small oscillations may falsely suggest the conclusion of a warm-up phase, it is essential to smooth the data. This smoothing process helps mitigate the influence of high-frequency noise and enables a clearer analysis of the overall behavior.

Outlier detection in the \kbkssd is performed by dividing the time series into several subsets. For each subset, the median is calculated as a representative value. The individual data points within each subset are then compared against specified upper and lower percentiles to identify outliers. If any data point falls outside the designated percentile range, then it is considered an outlier.

When an outlier is detected, it is replaced by the median of its corresponding subset. This replacement effectively preserves the general trend of the time series while eliminating the influence of anomalous values. By employing this technique, the time series can be effectively smoothed, allowing for more accurate trend analysis and reducing the risk of drawing false conclusions from erroneous data points. 

\subsection{Detection of a warm-up phase}\label{sec:warmUpDetect}
After obtaining the smooth data, the discrete convolution given by

\begin{align}
    (a * t)_n = \lim_{i=\infty} \sum^i_{m=-i} a_i * t_{n-i}
\end{align}
is applied, with $t$ and $a$ denoting the time series we are detecting the steadiness of, and the applied convolution kernel, respectively.
Two different kernels are used, both based on
\begin{align}
    a_i = \begin{cases}
            1, \quad 1 \leq i \leq \frac{n_k}{2}\\
            -1, \quad \frac{n_k}{2} + 1 \leq i \leq n_k,
          \end{cases}
\end{align}
where $a_i$ denotes the $i$-th element of the kernel and $n_k$ denotes the kernel size.

A first ``large'' kernel $a_l$ is selected to cover the whole time series of length $n_t$, i.e. $n_t = n_k$, while the other ``short'' kernel  $a_s$ is of fixed length, short enough to be able to account for steps located at tails of an investigated time series. For our experimentation, a length of 15 samples was heuristically chosen, because in our context it strikes a balance between being small enough to capture localized changes, such as steps in the tails of the time series, and large enough to avoid being overly sensitive to noise. A smaller kernel is effective at detecting high-frequency features like sudden peaks or jumps, which are characteristic of steps in the tails. However, excessively small kernels may not provide sufficient context, potentially leading to misdetections\,\cite{xie2019modeling}. Once the kernel size is reduced beyond a certain point, further decreases likely offer diminishing returns, as the kernel continues to capture the key features without losing accuracy. Furthermore, studies suggest that smaller kernel sizes are often preferable for time series analysis, as they focus on local features and reduce computational complexity\,\cite{he2020omni}. 

\begin{figure}
    \centering
    \includegraphics[width=\linewidth]{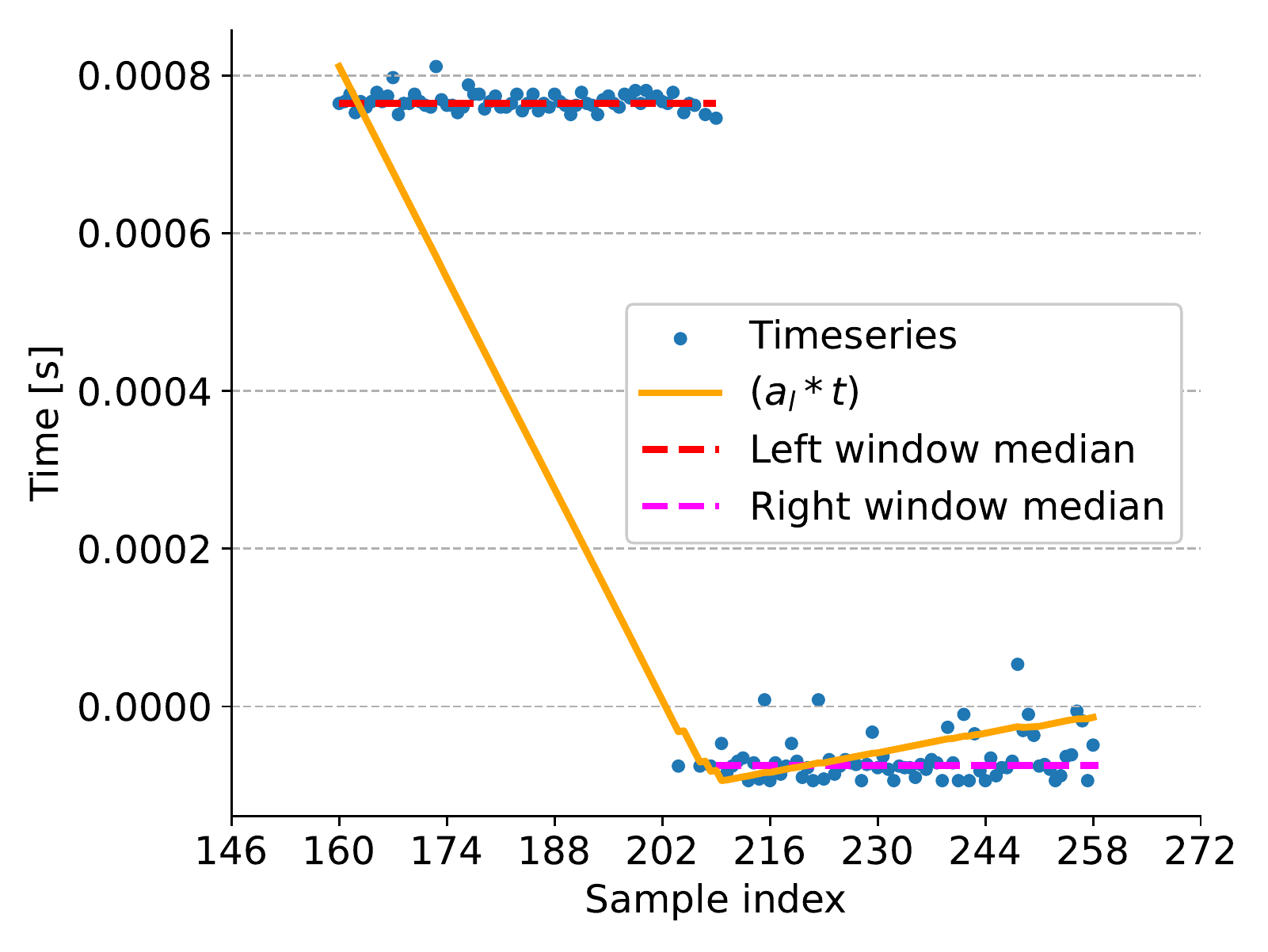}
    \caption{The illustration of step detection via convolution with a large kernel.}
    \label{fig:conv}
\end{figure}

After the convolution, the minima of both the resulting sequences $(a_l * t)$ and $(a_s * t)$ are located, thus detecting a step-down, if present. This principle, containing a large kernel and the windows with medians around its minimum, is illustrated in \Cref{fig:conv}.

After localization of both steps, it is decided which one of them, if any, is significant enough, based on the median comparison with its surroundings. The process is explained via the combination of an activity diagram and a pseudocode in the appendix.

\subsection{Kelly's Steady-State Detection}
After the potential detection of the warm-up phase termination, as it will be described in \cref{sec:warmUpDetect}, the subsequent analysis of the time series is performed using \kssd approach \cite{kelly2013steady}. Originally developed for the real-time monitoring of chemical reactors, this method enables the identification of steady-state conditions, which are characterized by minimal fluctuation in system parameters. Kelly’s approach is particularly advantageous as it not only facilitates the retrospective analysis of completed time series but also supports ongoing, dynamic monitoring of the system’s behavior in an ``online'' fashion. This capability allows for continuous assessment of whether the system has stabilized or is undergoing transient changes, providing a valuable tool for real-time process control. 

In summary, the \kssd algorithm combines statistical methods and mathematical expressions to effectively detect steady states in continuous processes, ensuring that the process is operating within acceptable limits for optimization and control.

The first assumption of the \kssd algorithm is that the time series behavior is operating with a non-zero slope multiplied by its relative time within the window given by
\begin{align}
    x_t = mt + \mu a_t,
\end{align}
where $mt$ denotes the deterministic drift component; $\mu$ denotes the mean of a hypothetical stationary process equaling the sample mean over the time window with zero slope; $a_t$ denotes the independent and identically distributed random error series with zero mean and a standard deviation $\sigma_a$. The index $t$ denotes the iteration at which the sample is collected, thus representing a so-called ``random walk with drift''~\cite{box2015}. Without going into in-depth mathematical reasoning behind the method, it can be seen that the mean of $x_t$ denoted as $\mu$ is defined as:
\begin{align}
    \mu = \frac{1}{n} \left( \sum^n_{t=1} x^t - m \sum^n_{i=1}t \right)
\end{align}
\noindent and that the expected value of $a_t - a_{t-1}$ is zero with a standard deviation of $2\sigma_a$, which
can be estimated as
\begin{align}
    \sigma_a = \sqrt{\frac{1}{n-2}\sum^n_{t=1}(x_t - mt -\mu)^2}.
\end{align}
Now, together with a specified t-test threshold value $t_{\text{crit}}$ at a specific significance level, everything necessary is available to test the null hypothesis that the time series values are steady around $\mu$. This hypothesis can be formalized as:
\begin{align}
    \begin{cases}
        y_t = 1, \quad |x_t - \mu| \leq t_{\text{crit}}\sigma_a\\
        y_t = 0.
    \end{cases}
\end{align}
The likelihood that the null hypothesis is false can be expressed by
\begin{align}
    \mathcal{L}(y) = \frac{1}{n}\sum^n_{t=1}y_t,
\end{align}
as it is the fraction of time where the time series is considered to be steady.

\subsection{Parameter Sensitivity Analysis}\label{sec:sensitivityAnalysis}
The sensitivity analysis of \kbkssd algorithm is performed via Sobol's indices~\cite{saltelli2010variance}, as the model's response is not generally linear and one-at-a-time analysis is not sufficient due to the presence of significant interactions of higher orders w.r.t. the model parameters.

Sobol's indices are used in global sensitivity analysis to quantify the contribution of each input variable to the output variance of a mathematical model. Given a model \( f \) with input variables \( \mathbf{X} = (X_1, X_2, \dots, X_d) \) and output \( Y = f(\mathbf{X}) \), the total variance of the output \( Y \) is given by
\begin{align}
\text{Var}(Y) = \text{Var}(f(\mathbf{X})),
\end{align}
where
\begin{align}
    f(\mathbf{X}) &= \sqrt{\sum_{i=1}^d |X_i|^2}
\end{align}
represents L2-norm, i.e. the sum of squared errors cost function, which quantifies the deviation from a manually-selected ground truth.

Sobol's first-order sensitivity indices \( S_i \) measure the contribution of each individual input variable \( X_i \) to the total variance of the model output. Mathematically, the first-order index for \( X_i \) is defined as the ratio of the variance of the output conditioned on \( X_i \) to the total variance of the output given by
\begin{align}
S_i = \frac{\text{Var}_{X_i}(f(\mathbf{X}))}{\text{Var}(f(\mathbf{X}))},
\end{align}
where \( \text{Var}_{X_i}(f(\mathbf{X})) \) represents the variance of \( f(\mathbf{X}) \) when all inputs except \( X_i \) are held constant.

In addition to first-order indices, Sobol's method includes higher-order sensitivity indices, which measure the contribution of interactions between input variables to the total variance. For example, the second-order index \( S_{ij} \) represents the contribution of the interaction between variables \( X_i \) and \( X_j \), and is given by
\begin{align}
S_{ij} = \frac{\text{Var}_{X_i, X_j}(f(\mathbf{X}))}{\text{Var}(f(\mathbf{X}))},
\end{align}
where \( \text{Var}_{X_i, X_j}(f(\mathbf{X})) \) denotes the variance of the output due to the interaction between \( X_i \) and \( X_j \). Additionally, Sobol’s total sensitivity indices \( S_i^{\text{tot}} \) measure the total effect of \( X_i \), which includes both the direct effect of \( X_i \) and its interactions with other inputs. The total index is defined as:
\begin{align}
S_i^{\text{tot}} = 1 - \frac{\text{Var}_{X_i^\perp}(f(\mathbf{X}))}{\text{Var}(f(\mathbf{X}))},
\end{align}
where \( X_i^\perp \) represents all input variables excluding \( X_i \). This total index captures the influence of \( X_i \) on the output, both directly and through interactions with other variables.

In this paper, SALib implementation\footnote{\url{https://salib.readthedocs.io/en/latest/}} of Sobol's sensitivity analysis was used.

\subsection{Model Fitting}\label{sec:modelFit}
The model optimal configuration was approximated via a grid search over 15000 different parameter configurations w.r.t. the Manhattan distance
\begin{align}
    m(\mathbf{X} | \theta) = \sum^n_{i=1} \left|p(\mathbf{X} | \theta) - g(\mathbf{X} | \theta) \right|,\label{eq:manhattanDist}
\end{align}
with $p$ and $g$ denoting the model's prediction and the ground-truth value corresponding to the vector of input variables $\mathbf{X}$ representing the time series included in the ground truth and the selected set of parameters $\theta$. The time series evaluated as unsteady by \kbkssd were not considered in the ``training'' process. The process itself was straightforward, selecting the set of parameters $\tilde{\theta}$, for which the Manhattan distance from the ground truth is minimal, as given by
\begin{align}
    \min_{\theta} m(\mathbf{X} | \theta).
\end{align}
 \section{Results}
\label{sec:results}
The reference set that represents our ground truth was obtained in three phases: (i) random selection of 586 of time series from a dataset\footnote{\url{https://github.com/SEALABQualityGroup/steady-state}}, (ii) binary classification of (un)steadiness of every time series, and (iii) manual selection of index where the steady state starts in time series judged as steady. The binary classification in step (ii) was performed via \textit{judgment aggregation}~\cite{yager1994weighted,yager1998fusion,nehring2022median} in a group of five researchers. The steady-state index was then selected individually by every researcher, and the five indices were checked for clusters with DBSCAN method~\cite{khan2014dbscan}. If there were three or five indices clustered, the middle point was selected as a reference steady state index. We did the same where all indices were scattered, i.e. where the ``judges'' more heavily disagreed. Finally, if there were four indices clustered, the third of them is taken as a reference point, as the more conservative judgment. We remark that, besides other analyses, we have also separately analyzed the cases of agreement and disagreement. From now on we will refer to this set as \jagt.

The objective of this section is to address the research questions, introduced in \Cref{sec:intro}, by conducting a comprehensive evaluation of \kbkssd algorithm in two key aspects. First, we perform a comparative analysis between \kbkssd and \cpssd, focusing on their respective accuracy and reliability to detect steady state. Second, we investigate the sensitivity of \kbkssd to changes in its key parameters, aiming to quantify how changes in these parameters influence its overall performance and robustness. This dual approach provides a thorough understanding of both the comparative efficacy and the parameter sensitivity of \kbkssd. 

\subsection{RQ1. How does KB-KSSD compare to existing
methods in terms of the accuracy of steady-state detection?}
In this section we describe a thorough numerical analysis of different aspects of \kbkssd, providing both an in-depth comparison with the previous S-o-A approach and Sobol's sensitivity analysis describing how ``influential'' model's parameters are for its predictions.

The optimal configuration, determined through this grid search described in \Cref{sec:modelFit}, consists of a window size of 100 samples for outlier removal, a window size of 500 samples to assess steadiness as per \kbkssd, a critical t-test value of 4.0 within Kelly’s steadiness evaluation, a window size of 70 samples for step detection, and a probability threshold of 0.95 for determining steadiness according to Kelly's part again.

\subsubsection[a)]{How good the methods are at classifying time series
as steady or non-steady?}
\begin{figure}
    \centering
    \includegraphics[width=\linewidth]{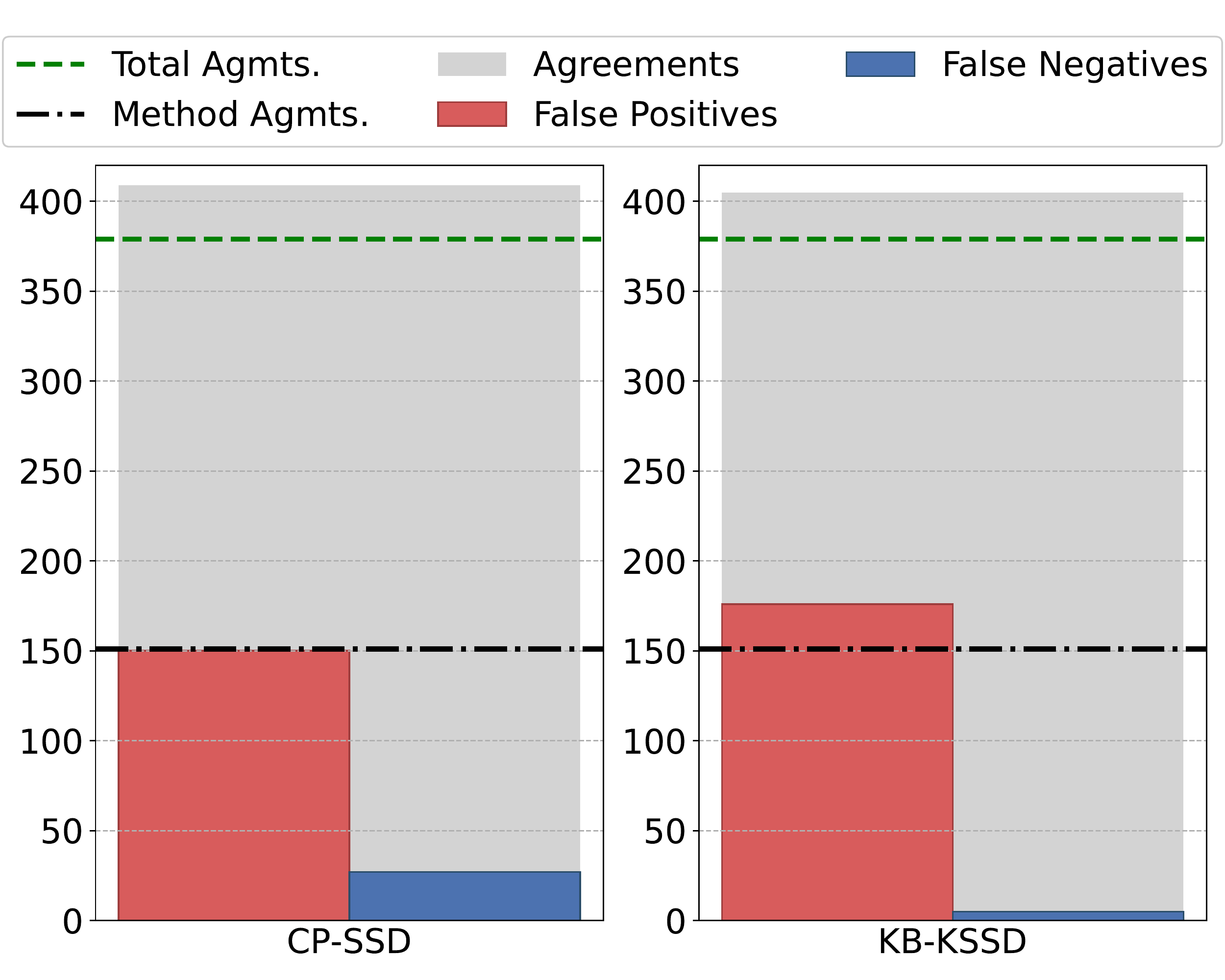}
    \caption{Number of (dis)agreements w.r.t. the larger dataset including manually-detected unsteady series. The number of total agreements denotes situations where both methods and the ground truth agree with each other, while the number of method agreements denotes the situation where both methods agree with each other, but not with the ground truth.}
    \label{fig:no_agreements}
\end{figure}

Firstly, in \Cref{fig:no_agreements} we present the number of agreements and disagreements regarding the (un)steadiness classification for both \kbkssd, \cpssd, and \jagt  including manually-selected unsteady time series. There were 586 time series considered in the ``full'' ground truth set, 407 of them were labeled as steady. Both \kbkssd and \cpssd were evaluated on all the time series, and it was found that both methods performed very similarly, achieving 409 and 405 agreements, i.e. $\sim$69.8\% and $\sim$69.1\%, w.r.t. \cpssd and \kbkssd, respectively. \kbkssd classified 176\,($\sim$30\%) as steady, while they were not, as compared to 150\,($\sim$26\%) false positives obtained via \cpssd. For the false negatives (i.e. classifying as unsteady a time series that was labeled as steady in the ground truth), \kbkssd misclassified 5\,($\sim$0.86\%), while \cpssd 27\,($\sim$4.6\%). Both methods simultaneously agreed with the ground truth in 379\,(65.7\%) cases. Both methods were found to agree with each other and to disagree with the ground truth at the same time in 151\,(25.8\%) cases.

\begin{figure}
    \centering
    \includegraphics[width=\linewidth]{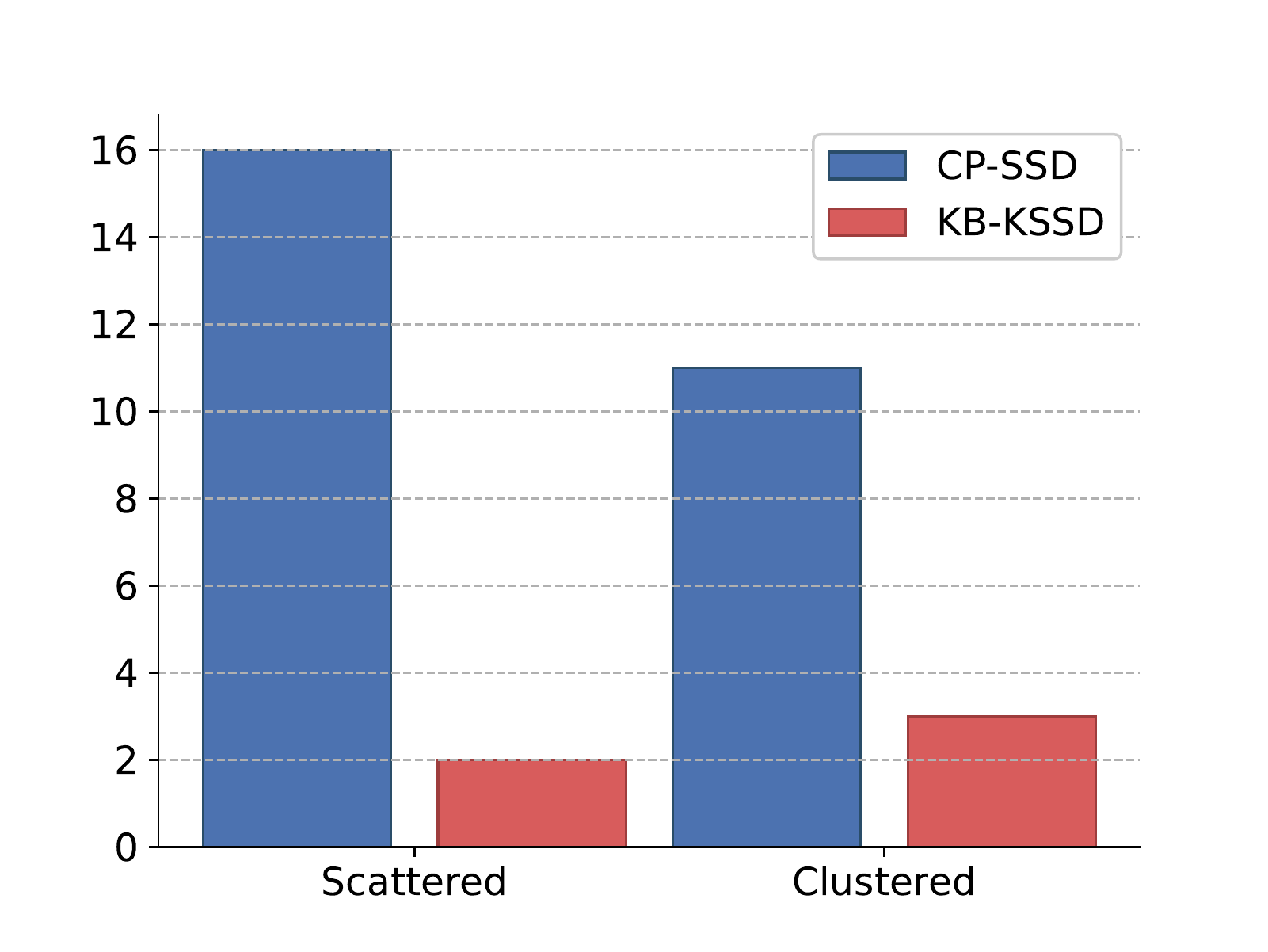}
    \caption{Number of detected unsteady time series by both approaches w.r.t. the ground truth labels.}
    \label{fig:no_unsteady}
\end{figure}

Furthermore, the number of false positives was partitioned w.r.t. the character of the ground truth labels, which can be: (i) clustered, in case of a strong agreement among human classifiers, or (ii) scattered, in case of some disagreement among human classifiers, coherently with the cases mentioned at the beginning of this section. In \Cref{fig:no_unsteady}, it is shown that \cpssd delivered falsely negative results in 16 and 11 cases for time series with scattered and clustered ground truth steadiness indices, respectively, while \kbkssd provided 2 and 3 of them.

\begin{rqbox}
Both methods are similarly efficient when it comes to the binary classification of (un)steadiness. Even if \kbkssd tends to classify a time series incorrectly as steady more often than \cpssd, \kbkssd is more reliable with respect to false negatives, \ie classifying the series as unsteady, while it is steady.
\end{rqbox}

\subsubsection{How distant (in terms of iterations and/or time) the two methods are from the ground truth?}
For obvious reasons, the following investigation on the ability to correctly find the index of steady-state beginnings is only based on the set of steady time series as reference ground truth. 

Consequently, we analyzed both the raw differences from the ground truth and their absolute values to correctly assess a possible directional bias in model results, \kbkssd reliability, and accuracy. All the statistical tests were performed with a 0.05 level of significance.

\begin{figure}
    \centering
\includegraphics[width=\linewidth]{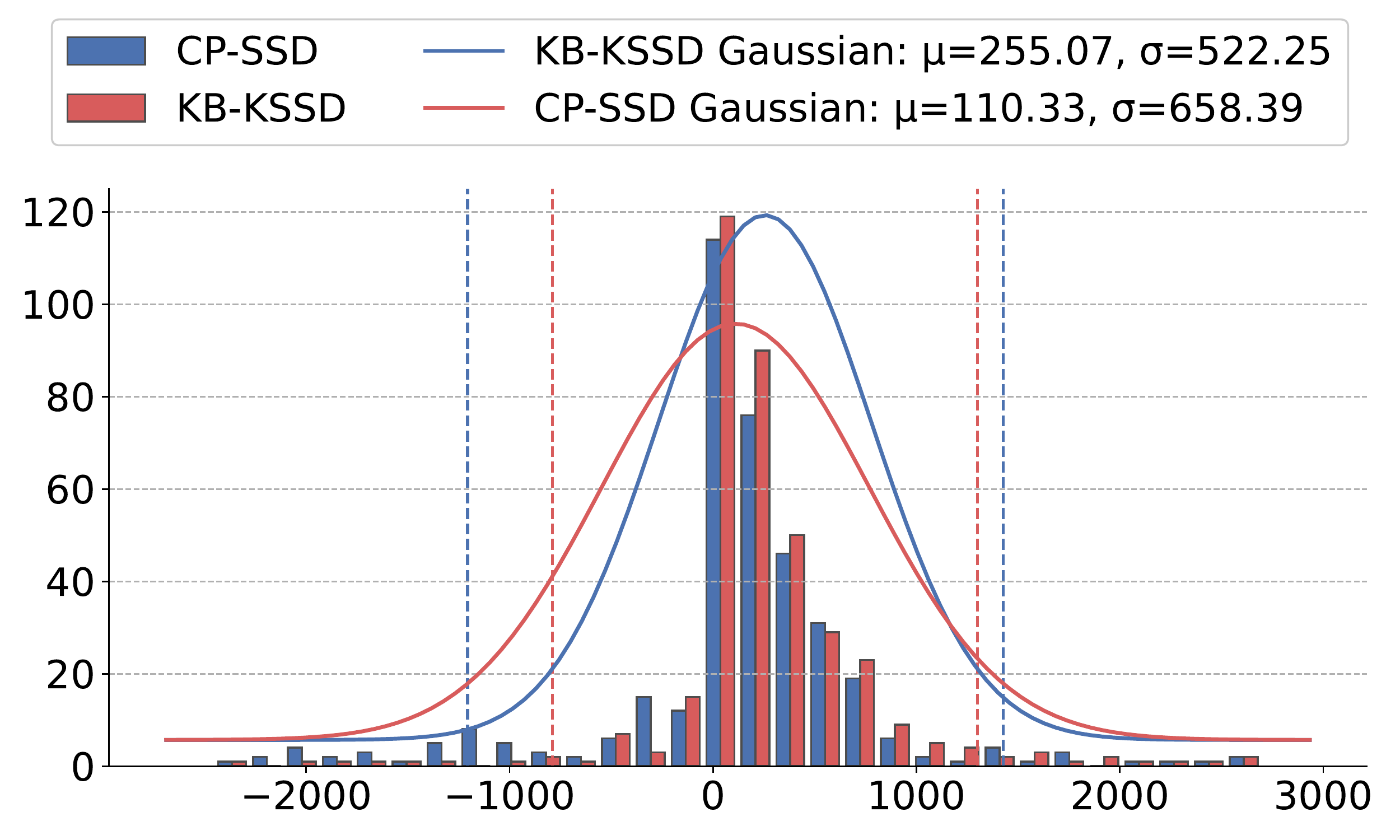}
    \caption{Distribution of detection differences (\cpssd - \kbkssdns) for all time series. The plot also visualizes the range, where 95\% of all errors stay within the vertical lines and there are Gaussians included to visualize an approximate normal distribution with the same $\mu$ and $\sigma$ to the real error distributions for easier readability.}
    \label{fig:diffs_all}
\end{figure}
\begin{figure}
    \centering
    \includegraphics[width=\linewidth]{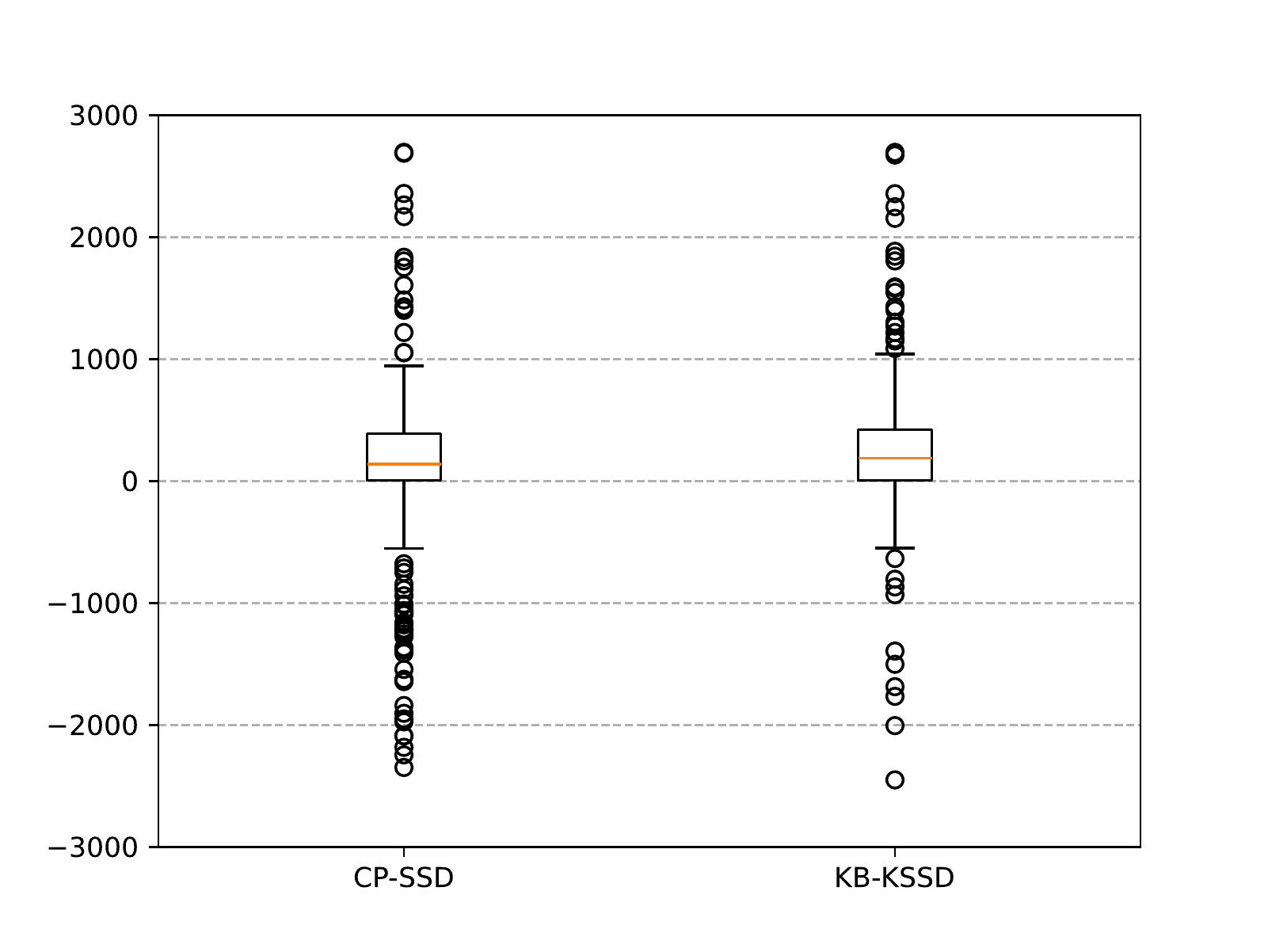}
    \caption{Detection differences (\cpssd - \kbkssdns) for all time series.}
    \label{fig:box_diffs_all}
\end{figure}

\begin{figure}
    \centering
    \includegraphics[width=\linewidth]{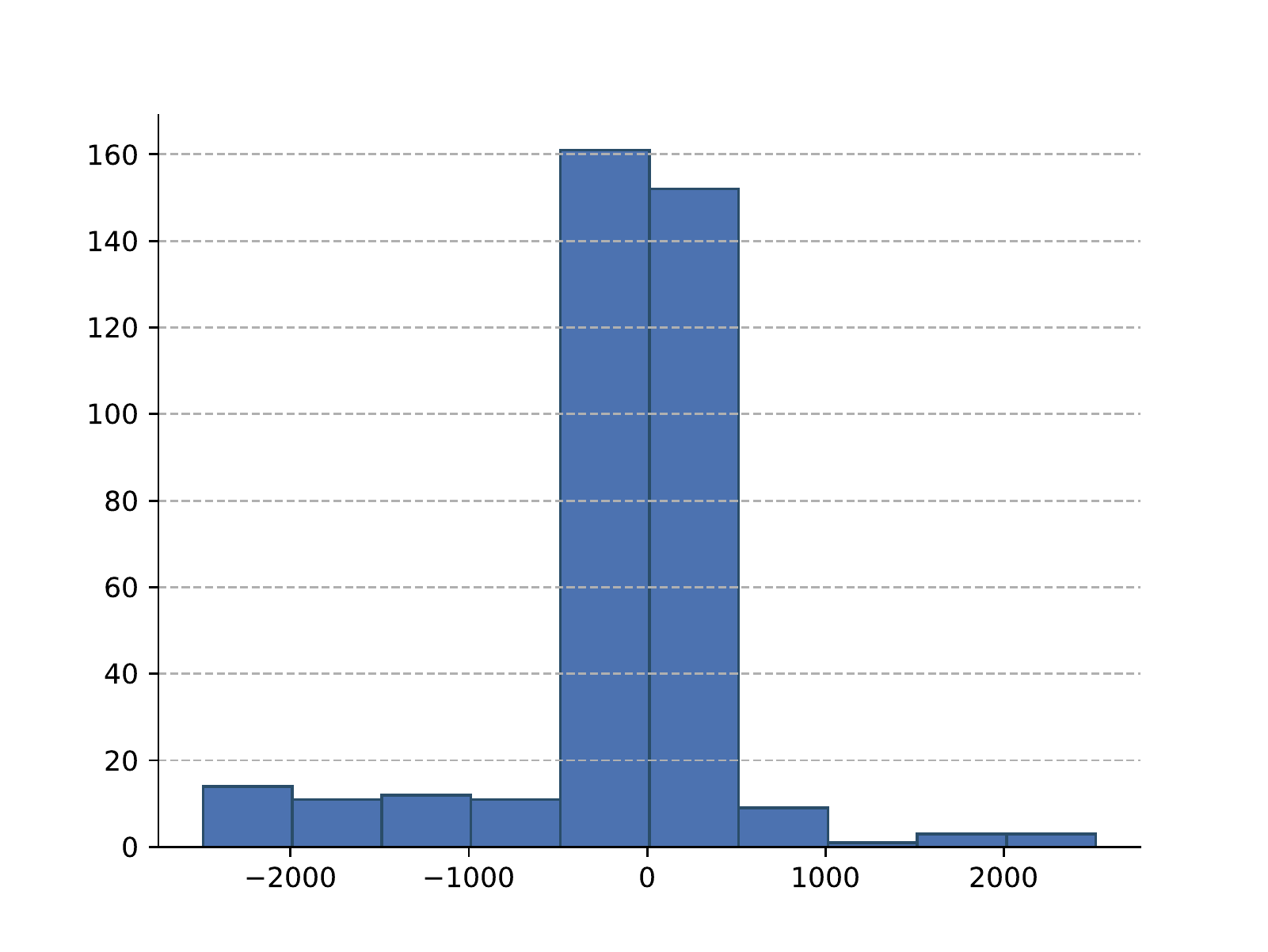}
    \caption{Differences between paired raw errors.}
    \label{fig:raw_err_diffs}
\end{figure}

By analyzing the raw errors (i.e., the differences between the ground truth and the value identified via \kbkssd) visualized in \Cref{fig:diffs_all,fig:box_diffs_all}, we calculated the mean raw errors and the corresponding standard deviations for \kbkssd and \cpssd as $\mu_N = 255.7$, $\sigma_N = 522.3$ and $\mu_R = 110.3$, $\sigma_R = 658.40$, respectively. Based on these findings we can assert that 95\% of all errors fall within the intervals $\langle -789.4; 1299.6 \rangle$ and $\langle -1206.5; 1427.1 \rangle$ for \kbkssd and \cpssd, respectively. As it can be seen, the 95\% interval is longer by $\sim$26.1\% in the case of \cpssd. Also, the mean raw errors of both methods are larger than zero, thus implying a directional bias in both of them. Since \kbkssd delivers a higher mean, we tested whether the difference in bias is statistically significant. Hence, we checked the differences between the sets of errors, shown in \Cref{fig:raw_err_diffs}, to assess their normality and symmetry, thus testing assumptions for t-test\,\cite{david1997paired} and Wilcoxon signed-rank test\,\cite{conover1999practical}. As the Shapiro-Wilk test\,\cite{shapiro1965analysis} resulted in statistic $\sim$0.72 and p-value 0, it can be safely concluded that the distribution of differences is not normal, rendering the t-test unfit for making the decision. For the symmetry assessment, we computed the skewness and we obtained a -1.14 value, which signifies a moderate skewness to the left, rendering the Wilcoxon test also unreliable\,\cite{bulmer2012principles}. With this in mind, we settled for a sign test, thus observing that the difference in directional bias is significant, i.e. \kbkssd tends to ``underestimate'' the position of the steady state more than \cpssd.
Albeit not so large considering the whole difference range, this remains one of the \kbkssd properties, which needs to be further addressed in future work.
To be sure that the models' errors do not follow the same distribution, we checked it with Cramer-von Mieses test\,\cite{vonmises1931distribution} to uniformly evaluate the differences, Kolmogorov-Smirnov test\,\cite{smirnov1948table} to evaluate the differences emphasizing the behavior around the expected value, and Anderson-Darling test\,\cite{anderson1952asymptotic}, mainly focused on distribution tails. All the tests agreed that the raw error distributions are different.

Thereafter, we intended to assess the models' reliability, i.e. the ability to provide reasonable results without many significant outliers. In particular, we wanted to test if the difference in the distributions' variances is significant, or if the previous results were mostly influenced by the difference in the expected values. For this goal, we utilized Levene test~\cite{levene1960robust,brown1974robust}, which has rejected the possibility of similar variances, thus confirming that \kbkssd provides more stable behavior.

\begin{figure*}
    \centering
    \begin{subfigure}[t]{0.45\textwidth}
        \centering
        \includegraphics[width=\linewidth]{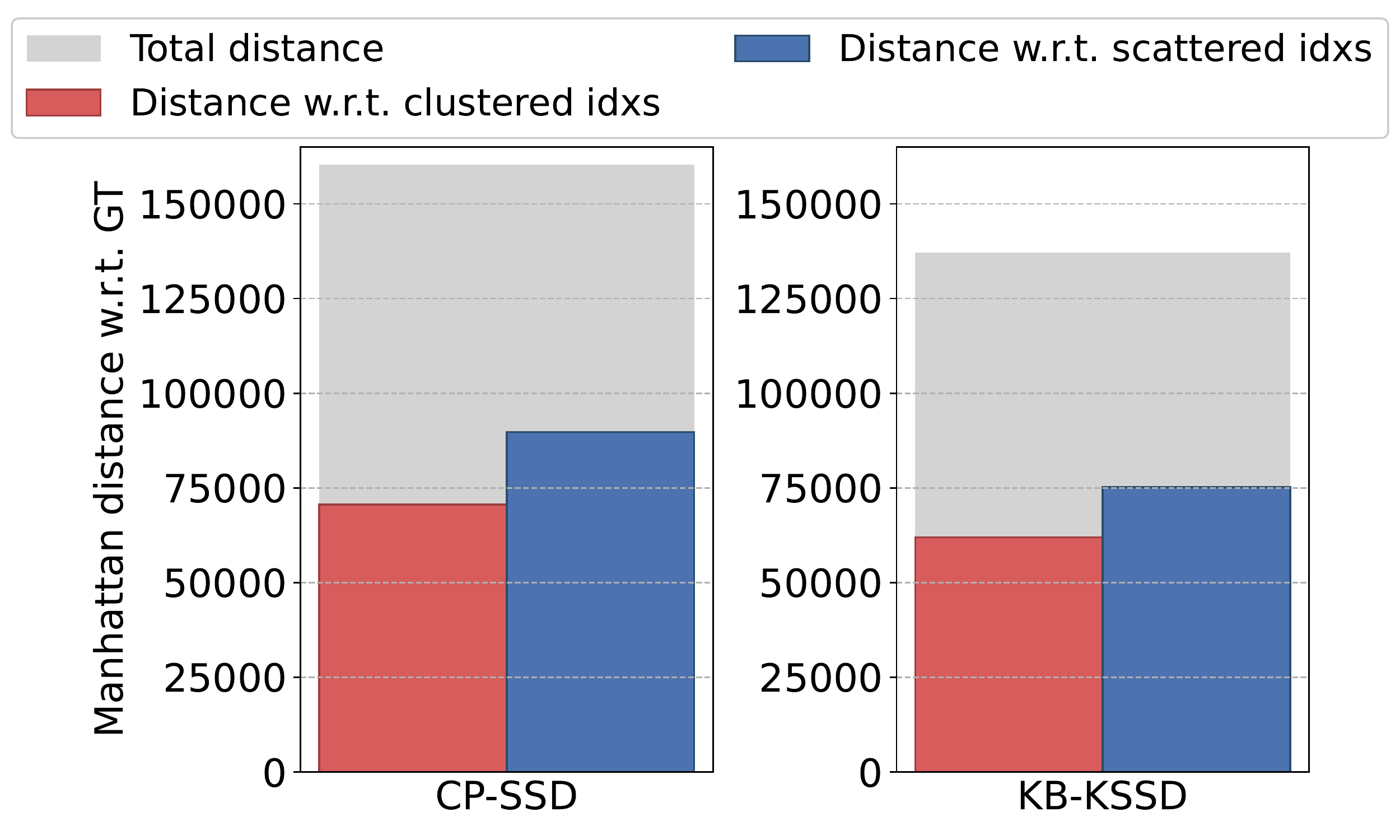}
        \caption{Bar plot of Manhattan distances from the ground truth of \cpssd and \kbkssd}
        \label{fig:manhattanDist}
    \end{subfigure}
    \hfill
    \begin{subfigure}[t]{0.45\textwidth}
    \centering
    \includegraphics[width=\linewidth]{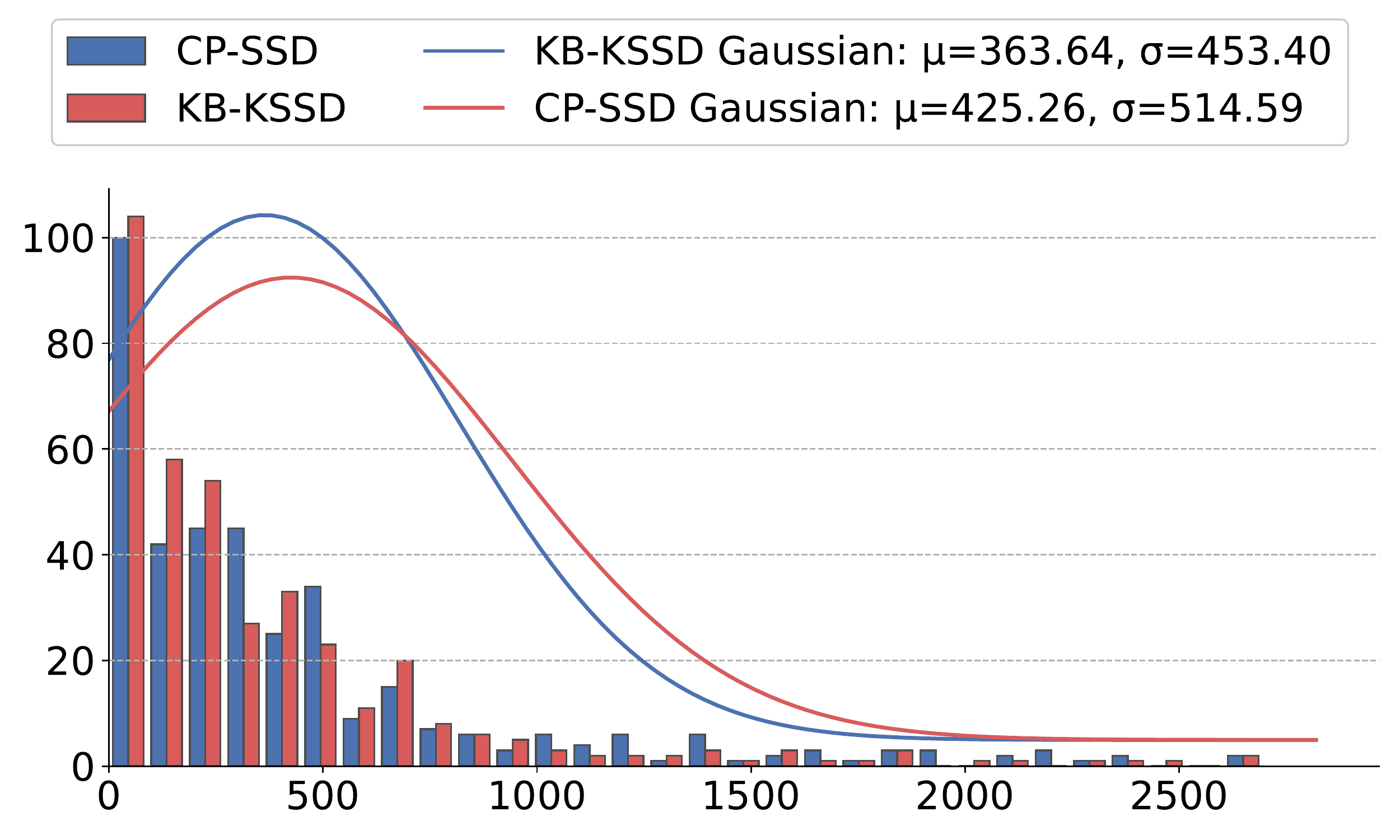}
    \caption{Histogram of Manhattan distances from the ground truth.}
    \label{fig:manhattanDistHist}
    \end{subfigure}
    \caption{Manhattan distances from the ground truth.}
    \label{fig:mDistance}
\end{figure*}

To mitigate the possible problem with false discoveries, i.e. falsely rejecting the null hypothesis, we applied Benjamini-Hochberg correction~\cite{benjamini1995controlling}. All results of these tests, i.e. the significance of differences in both expectation values and variances, can be seen in detail in \Cref{tab:rawStats}.
\begin{table}
\centering
\begin{tabular}{lccc}
\toprule
Test & Statistic & p-value & BH-adjusted p-value \\ 
\midrule
Sign   & 0.58 & 0    & 0.01749295 \\ 
CvM    & 0.51 & 0.04 & 0.02313882 \\ 
KS     & 0.11 & 0.02 & 0.02150996 \\ 
AD     & 3.61 & 0.01 & 0.03670551 \\ 
Levene & 6.21 & 0.01 & 0.02150996 \\ 
\bottomrule
\end{tabular}
\vspace{.5em}
\caption{Statistical tests comparing the raw error distributions.}
\label{tab:rawStats}
\end{table}
The accuracy of both models was established based on their Manhattan distance, given by \Cref{eq:manhattanDist}, from the ground truth. As visualized in \Cref{fig:manhattanDist,fig:manhattanDistHist}, we found out that \kbkssd performed with a total error of 137094, while \cpssd performed with 160323, i.e. with an error larger by $\sim$14.5\% than the \kbkssd one. Considering the accuracy for the clustered and scattered indices in the ground truth separately, \kbkssd performs with the errors of 61924 and 75170, respectively, while \cpssd performs with 70579 and 89744. It can be seen that, as expected, while both methods perform better when the indices were clustered (i.e., when the steady state was identified by human classifiers in agreement), \kbkssd is more accurate in both cases, with \cpssd exceeding its errors by $\sim$12.3\% and  16.2\%. About the expected performance, the mean absolute values and the standard deviations are $\mu_{NA} = 363.6$, $\sigma_{NA} = 453.4$ and $\mu_{RA} = 425.3$, $\sigma_{NA} = 514.6$ for \kbkssd and \cpssd, respectively. Thus, considering the error magnitudes, \cpssd performs with an error higher by $\sim$14.5\% on average, while the standard deviation of its absolute errors is higher by $\sim$11.9\%, thus signifying more frequent errors by large magnitude.

To further scrutinize these observations and to generalize them to the assumed population instead of only the limited set of observations, we checked again whether the differences in the populations' expected values were significant.

\begin{figure}
    \centering
    \includegraphics[width=\linewidth]{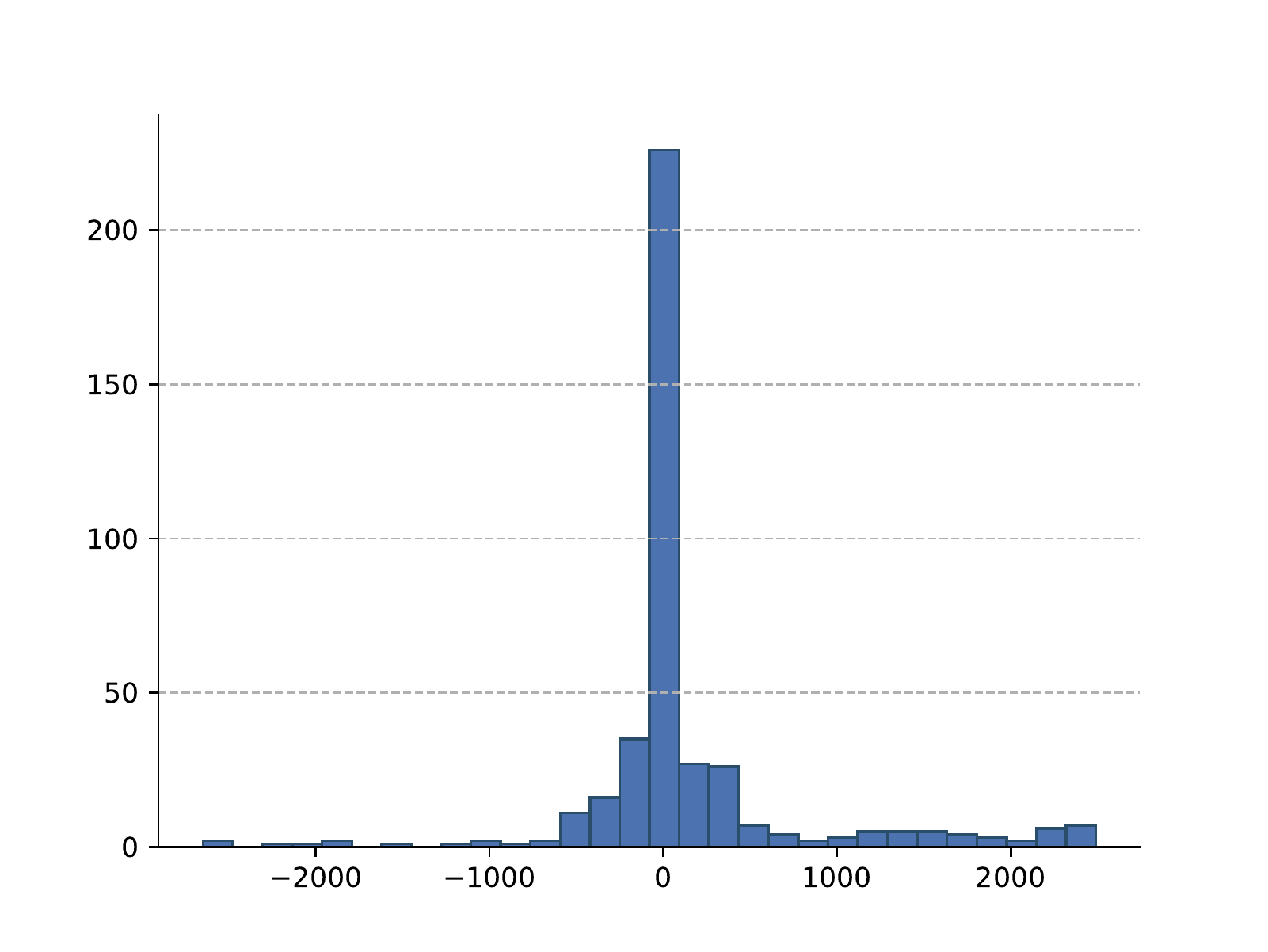}
    \caption{Histogram of differences of Manhattan distances from the ground truth.}
    \label{fig:manhattanDiffsHist}
\end{figure}

For this, we needed to select the convenient statistical test again. The differences between absolute errors, visualized in \Cref{fig:manhattanDiffsHist}, are not normal in the sense of Shapiro-Wilk, as shown by statistic $\sim$0.73 and a p-value 0. Thus, we need to discard the t-test. Moreover, the differences are not even symmetric, as the skewness equal to $\sim$0.59 suggests that the differences are skewed to the right rather significantly, thus rendering also Wilcoxon signed-rank test unreliable. The finally adopted signed rank test eventually suggested that neither of these absolute error distributions do possess a similar expected value with its statistic 0.67 and a p-value almost equal to 0.

The comparative analysis has shown that \kbkssd consistently outperforms \cpssd in several critical areas. \kbkssd demonstrated a lower false negative rate, classifying significantly fewer unsteady time series as steady. This shows that \kbkssd is more sensitive to unsteady periods, an essential feature for accurate time series classification. When detecting the onset of steady states, \kbkssd provided more precise results, with narrower error margins and more stable performance. While both methods exhibited some directional bias, \kbkssd showed fewer extreme outliers, further supported by the fact that \kbkssd had a narrower 95\% confidence interval for detection errors, indicating more reliable predictions. In terms of accuracy, \kbkssd reduced the total error (measured by Manhattan distance) by approximately 14.5\% compared to \cpssd. This performance improvement was evident in both well-clustered and more scattered ground truth data. These findings suggest that \kbkssd is not only more accurate in general, but also more capable of handling challenging or ambiguous scenarios, making it a better fit for real-world applications. Its reduced error magnitude and higher precision across different time series make it a more reliable and efficient tool for \ssdns.

\begin{rqbox}
\kbkssd outperforms \cpssd in terms of accuracy (error reduction by 14.5\%), reliability (fewer extreme outliers), and sensitivity (lower false negative rate). It provides more precise and stable results, with narrower error margins and a more reliable performance across different time series.
\end{rqbox}

\subsection{RQ2. How does KB-KSSD effectiveness vary while
modifying its main parameters?}
To address RQ2, \kbkssd was analyzed using Sobol's indices described in~\Cref{sec:sensitivityAnalysis}, thus assessing its sensitivity w.r.t.  changes in its parameters. As parameters to be investigated, we chose the size of the window used in \kssd (\texttt{prob\_win\_size} $\in \langle 400; 600 \rangle$), the size of the window for step-detection (\texttt{step\_win\_size} $\in \langle 50; 90 \rangle$), the critical value for the t-test included in \kssd (\texttt{t\_crit} $\in \langle 3; 5 \rangle$), the probability threshold for steadiness acceptance (\texttt{prob\_threshold} $\in \langle 0.75; 0.95 \rangle$), and the size of the window used for detecting outliers (\texttt{outlier\_win\_size} $\in \langle80; 120\rangle$). The investigated intervals were selected around the ``optimal'' configuration found by the grid search described in \Cref{sec:modelFit}. Moreover, there were 24576 samples generated for the analysis. 

These parameters were selected so as to capture most of the degrees of freedom, i.e. to represent most of the flexibility of \kbkssd. Other parameters, like percentiles in the outlier removal phase described in \Cref{sec:smoothing}, were not included, as they are not a core part of the method. Also, the size of the lesser kernel was not investigated in this work, as it is usually enough to select a small enough kernel to detect patterns in series tails, and from that point on the behavior does not change. 
\begin{table}[ht]
\footnotesize
\begin{tabular}{lcc}
\toprule
                   & $S_1$                    & \textbf{$S_T$}   \\
\midrule
prob\_win\_size    & 0.19718534 (0.15, 0.24)  & \textbf{0.46349316} (0.41, 0.51) \\ 
step\_win\_size    & 0.00631332 (-0.0, 0.01) & \textbf{0.01394031} (0.01, 0.02) \\ 
t\_crit            & 0.15670927 (0.11, 0.2) & \textbf{0.46206506} (0.41, 0.52) \\ 
prob\_threshold    & 0.1107751 (0.07, 0.15)  & \textbf{0.35747568} (0.31, 0.41) \\ 
outlier\_win\_size & 0.15383209 (0.12, 0.19)  & \textbf{0.26101527} (0.24, 0.28) \\ 
\bottomrule
\end{tabular}
\vspace{.5em}
\caption{Sobol's indices of the first order $S_1$ and the total ones $S_T$ together with corresponding confidence intervals for the 0.95 confidence level.}
\label{tab:sobol}
\end{table}

In this analysis, indices $S_1$, $S_2$, and $S_T$ denoting different sensitivity measures were considered. $S_1$ (first-order Sobol index) quantifies the direct effect of an input variable on the output variance, $S_2$ (second-order Sobol index) measures the contribution on the output of interactions between pairs of variables, and $S_T$ (total Sobol index) captures the total effect of an input, including its main effect and interactions with all other variables. $S_T$ is generally larger than or equal to $S_1$ as it accounts for both individual and interaction effects. In \Cref{tab:sobol} it can be seen that the most significant parameters w.r.t. their changes are \texttt{prob\_win\_size} and \texttt{t\_crit}, followed by \texttt{prob\_threshold}, \texttt{outlier\_win\_size} and \texttt{step\_win\_size}, while the model is almost ``immune'' to the change of \texttt{step\_win\_size} in the chosen interval. The first-order effects do maintain the influence order aside from the fact that the model is somewhat more sensitive to the change of \texttt{outlier\_win\_size} than to the change of \texttt{prob\_threshold}. The large confidence intervals and the fact that there are some badly converged (less than 0) indices of the second-order effects indicate that there were not enough samples to assess these separately in a reliable manner. All of the $S_2$ values are listed in a table in the appendix. That said, we can reasonably assume that \kbkssd is basically insensitive to changes in the interactions of \texttt{step\_win\_size} with \texttt{outlier\_win\_size}, \texttt{prob\_threshold} and \texttt{t\_crit} due to their indices being $\sim0$, with confidence intervals (-0.01, 0.02), (-0.01, 0.01) and (-0.01, 0.01), respectively. It can be seen that some negative values appear in the confidence intervals, which are due to lack of computation convergence. However, the estimations can still be considered reliable, as not only the estimated indices are almost zero, but the obtained intervals are also very narrow. It can be further assumed that the only significant interaction exists between \texttt{prob\_threshold} and \texttt{t\_crit}, as its $S_2$ index is $\sim 0.13$, i.e. the only one significantly higher than 0, and its confidence interval is (0.04, 0.23), i.e. the only one being fully above than 0. Still, an analysis with larger number of samples could allow to deduce some further information from $S_2$ indices, but it implies very high computation demands.

\begin{figure}
    \centering
\begin{subfigure}{0.5\columnwidth}
        \centering
        \includegraphics[width=\textwidth]{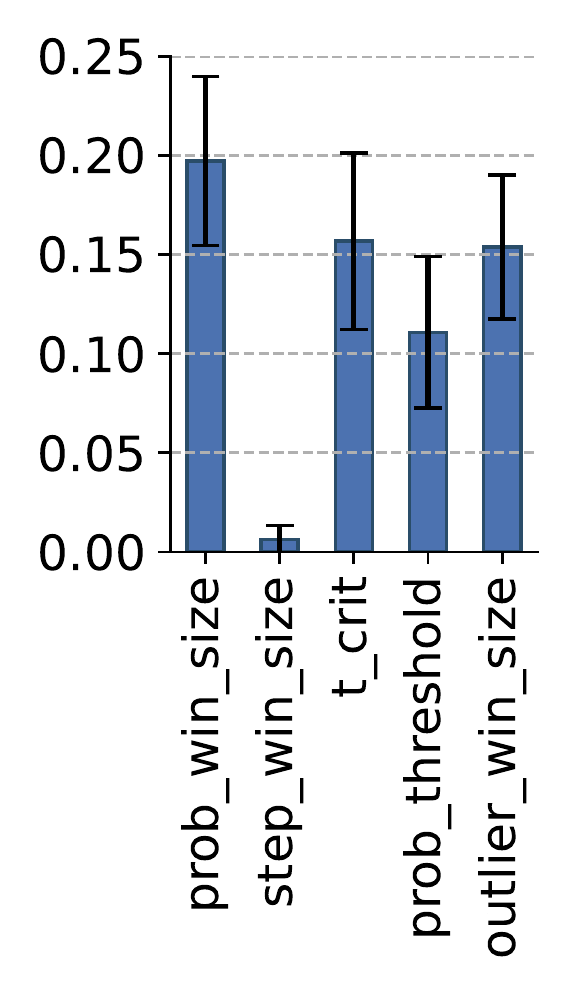}
        \caption{First order ($S_1$)}
    \end{subfigure}\begin{subfigure}{0.5\columnwidth}
        \centering
        \includegraphics[width=\textwidth]{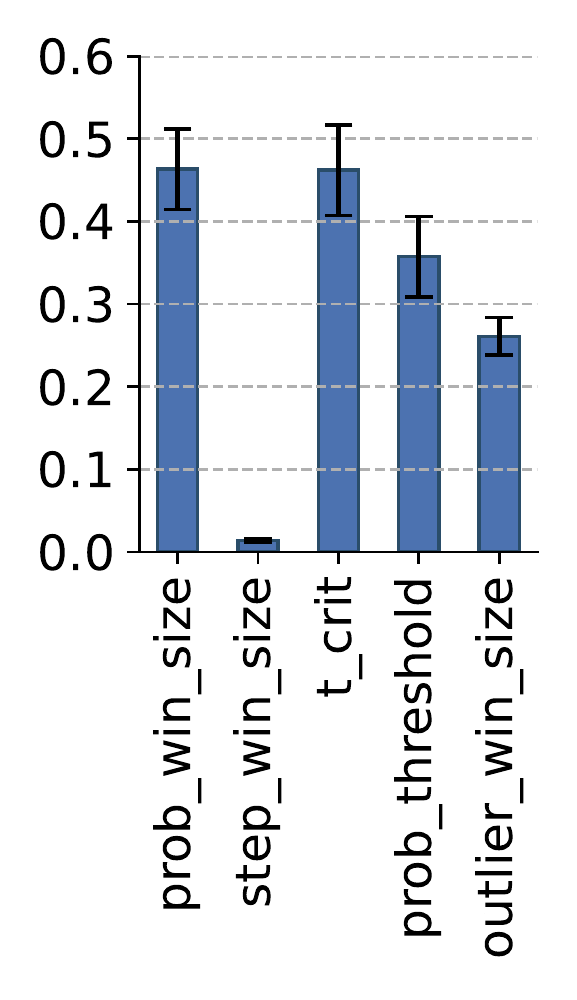}
        \caption{Total order ($S_T$)}
    \end{subfigure}
    
    \begin{subfigure}[t]{\columnwidth}
        \centering
        \includegraphics[width=\textwidth]{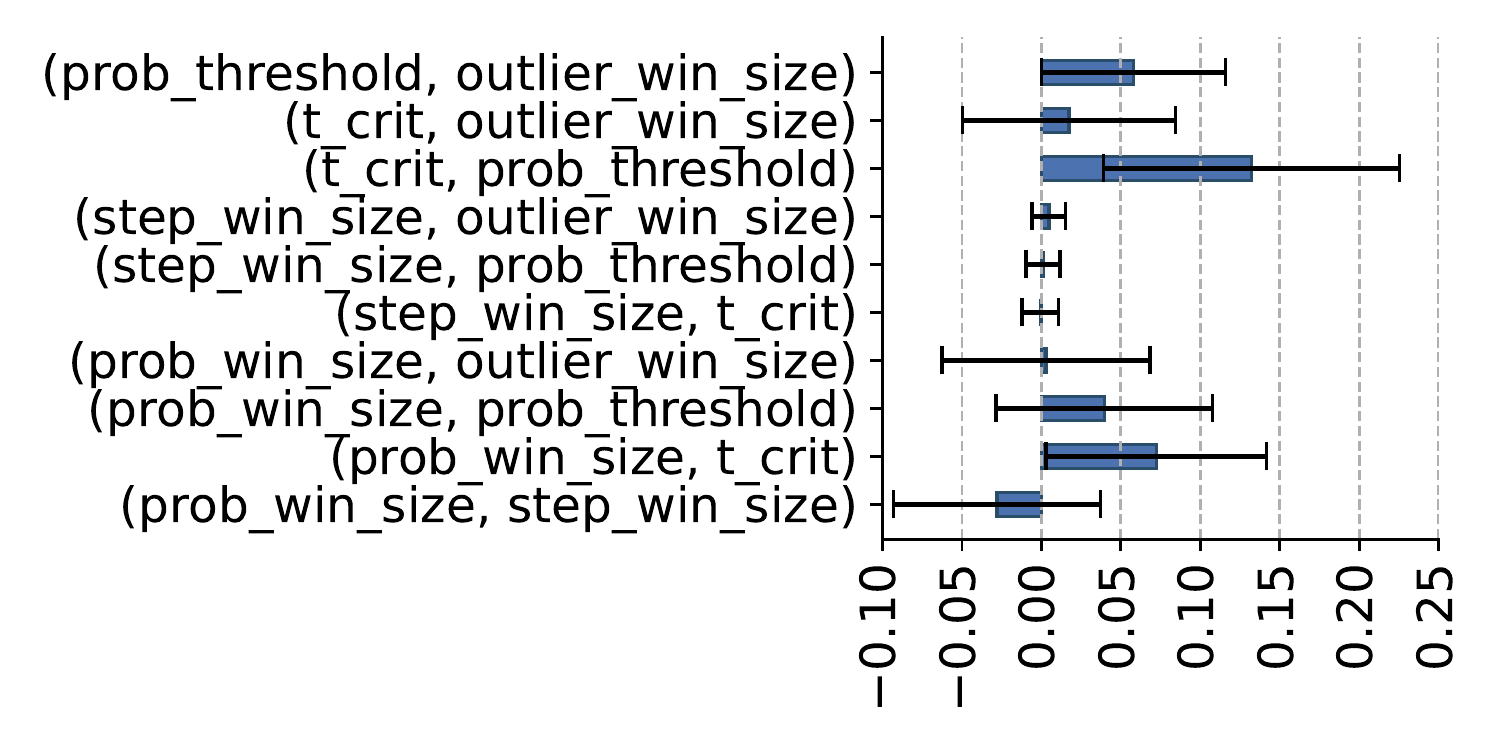}
        \caption{Second order ($S_2$)}
    \end{subfigure}
    \caption{Sobol's indices of the different selected parameters denoting the model's sensitivity w.r.t. them. The plot describes the indices of the 1st order $S_1$, 2nd order $S_2$, and the ``total'' index $S_T$ denoting the estimated behavior with all the higher order interactions involved. Also, the plot contains the confidence intervals of the index estimates for the 0.95 confidence level.}
    \label{fig:sobol}
\end{figure}

$S_1$, $S_2$ and $S_T$ are visualized together with their corresponding confidence intervals in \Cref{fig:sobol}. 
\begin{table}
\centering
\begin{tabular}{lcc}
\toprule
Parameters                                         & $S_T$                   & \textbf{$\delta$} \\ 
\midrule
prob\_win\_size    & 0.31653608 (0.26, 0.38) & \textbf{68.3\%}   \\ 
step\_win\_size    & 0.01159328 (0.01, 0.01) & \textbf{83.2\%}   \\ 
t\_crit            & 0.39023183 (0.31, 0.47) & \textbf{84.5\%}   \\ 
prob\_threshold    & 0.2987368 (0.22, 0.37)  & \textbf{83.6\%}   \\ 
outlier\_win\_size & 0.19648237 (0.16, 0.23) & \textbf{75.3\%}   \\ 
\bottomrule
\end{tabular}
\vspace{.5em}
\caption{Sobol's $S_T$ indices for the analysis performed with only 6144 samples and the $S_T$ percentage w.r.t. the $S_T$ obtained via 24576 samples (in \Cref{tab:sobol}). Performed at 0.95 significance level.}
\label{tab:sobolSmallN}
\end{table}
To assess the stability of $S_T$ estimation, another analysis with 6144 samples was performed, with the corresponding $S_T$ values listed in \Cref{tab:sobolSmallN}. In this case, the order of parameters obtained by considering the $S_T$ sensitivity is almost the same, with the model seemingly more sensitive to the change of \texttt{t\_crit} than to the change of \texttt{prob\_win\_size}. Talking about the confidence intervals, the analysis with a larger number of samples provides intervals with a mean $\sim$0.071 and the standard deviation of $\sim$0.04, while the analysis using a smaller number of samples provides intervals with mean of $\sim$0.102 and a standard deviation of $\sim$0.059, i.e. larger by $\sim$44.4\% and $\sim$47.5\%, respectively.

The sensitivity analysis of \kbkssd was conducted using Sobol's indices, focusing on key parameters such as window sizes and thresholds. It showed that \kbkssd is robust to variations in key parameters, with its performance largely determined by a few influential factors. Among the parameters studied, the size of the window used for steadiness evaluation (\texttt{prob\_win\_size}) and the critical value for the t-test (\texttt{t\_crit}) had the most significant impact on performance, while other parameters like \texttt{step\_win\_size} have minimal impact. Despite this sensitivity, the method maintained stable and reliable results across a range of parameter settings, indicating that it can be effectively tuned without risking large drops in performance.

\begin{rqbox}
\kbkssd is robust to variations in key parameters, with its performance largely determined by a few influential factors. The size of the window used for steadiness evaluation and the critical value for the t-test had the most significant impact on performance.
\end{rqbox}

 \section{Threats to validity}
\label{sec:threats}

\medskip
\paragraph*{Construct validity}
We assume that benchmarks reaching a steady state will do so within the defined execution time. Some benchmarks may require more time to stabilize. However, our selected dataset was constructed by running benchmarks significantly longer than usual (each benchmark was run at least $3\,000$ times across 10 forks)~\cite{TrainiCPT23}, exceeding typical developer practices and experiments in previous literature~\cite{Barrett2017}.

The process of establishing the ground truth relies on human judgment for both binary classification and steady-state index selection, which introduces potential subjectivity and inconsistency. While judgment aggregation was employed to mitigate individual biases, as done in previous studies \cite{yager1994weighted,yager1998fusion,nehring2022median}, the method does not eliminate the possibility of systematic bias among the five researchers. This might lead to a ground truth that does not fully represent the true underlying characteristics of the data, affecting the validity of subsequent analyses and conclusions.

\medskip
\paragraph*{Internal validity}
The process of identifying and replacing outliers may inadvertently distort the true characteristics of the time series data. By relying on median-based filtering within predefined percentile thresholds (\Cref{sec:smoothing}), genuine but extreme variations in benchmark performance could be misclassified as noise and removed. Nonetheless, filtering outliers in time series data is a common and essential preprocessing step to improve the accuracy of changepoint and trend detection~\cite{Blazquez-Garcia21}.

Since only 10\% of the available time series were annotated, there is a risk of selection bias. Even though a one-time series was chosen for each benchmark to ensure representation, this subset may not fully capture the variability present in the entire dataset.

The use of a graphical tool for visual inspection and annotation introduces a potential source of inaccuracy. Since the tool approximates the selected data point to the closest one to the user's mouse click, slight misalignments may occur, leading to minor errors in the ground truth annotations.

Changes in annotating standards over time could impact consistency, as annotators may unintentionally shift their interpretation of the annotation criteria throughout the annotation process. To avoid that this would introduce inconsistencies in the ground truth, recurring meetings were held to share concerns and realign the group on the basis of the ongoing annotation experience.

\medskip
\paragraph*{External validity}
The external validity threats of this work are primarily inherited from the dataset chosen for evaluation. Specifically, the dataset is limited to GitHub repositories, making it uncertain whether the findings generalize to other open-source hosting platforms or industrial software. Additionally, the dataset focuses solely on microbenchmarking using JMH. While JMH is widely adopted, the applicability of our findings to other benchmarking practices or settings remains unknown.

\medskip
\paragraph*{Conclusion validity}
\kssd assumes that the random error component of the time series is independently and identically distributed (i.i.d.). However, in real-world benchmarking scenarios, performance measurements often exhibit autocorrelation, where successive data points are not entirely independent due to ongoing optimization occurring at the JVM or operating system level. If this assumption is violated, the \kssd algorithm may misinterpret correlated variations as steady-state phases. Nevertheless, \kssd parameters, namely the sliding window size, can be tuned to mitigate these effects, allowing for some adaptation to real-world deviations from the i.i.d. assumption, as it can be seen in similar real-world applications, where we encounter ``structural breaks'' in series \cite{bai2003,gama2014}. Also, the i.i.d. assumption could be eased or completely removed, when an \textit{autoregressive model (ARM)}  would be incorporated into \kssd and the residuals from the ARM would be used in \kssd instead, as it is a quite common application in different fields dealing with discrete time series~\cite{box2015,hyndman2018}.
 \section{Conclusion}
\label{sec:conclusion}
This study introduced and evaluated \kbkssd for detecting steady states in time series data, and it has compared its performance to \cpssd. The evaluation addressed two research questions by assessing both the comparative accuracy of the methods and the sensitivity of \kbkssd to variations in key parameters.

Overall, \kbkssd offers several advantages over \cpssd. It is more accurate in detecting both steady and unsteady states, with a significantly lower false negative rate and reduced error magnitude. The method is also more robust, delivering reliable results even when key parameters are modified. This makes \kbkssd highly suitable for applications requiring precise and consistent steady-state detection, especially in benchmarking scenarios where performance must be both accurate and reproducible. Its ability to handle ambiguous cases and deliver stable results across parameter variations provides added flexibility for users, making it a preferrable choice compared to existing approaches.

The presented model exhibits several limitations that should be addressed in future work. Firstly, the training process utilized a simple cost function, which could be enhanced to account not only for the overall deviation from ground truth, but also for the number of false positives and false negatives, directional bias, error variance, and both mean and median absolute errors. These metrics could be weighted differently to optimize model performance. Furthermore, this multi-objective optimization would benefit from the application of a more robust optimization technique with a strong exploration component, such as particle swarm optimization or evolutionary algorithms, rather than relying on a simple grid search. However, it is important to note that these methods introduce their own challenges, including the risk of convergence to suboptimal local minima, vanishing gradients, barren plateaus (for gradient-based methods), and other issues.

The ``short'' convolution kernel $a_s$ was fixed to a size of 15 based on an educated guess. This choice has not been rigorously investigated or optimized, and future work should explore different kernel sizes to assess whether a more refined approach could improve sensitivity and detection accuracy. Systematically evaluating the impact of varying kernel sizes would provide more insight into the model robustness and effectiveness across different scenarios. Furthermore, the \kbkssd sensitivity to changes in $a_s$ length would be better understood via Sobol's indices.

Another limitation of the current model is its inability to detect upward steps, as it is exclusively designed to identify downward transitions in the data. This constraint limits the model applicability, particularly in scenarios where upward shifts in the system behavior may occur, such as during steady-state analysis of software runtime. To address this, future work should focus on extending the model to capture both upward and downward steps, thereby enhancing its ability to fully characterize the dynamics of the system under various conditions.

Finally, it would be advantageous to implement more targeted pre-processing techniques tailored to the specific characteristics of the time series. This aspect is closely linked to the broader issue of time series classification, which will be the focus of future research efforts.
 
\section*{Acknowledgements}
Martin Beseda was supported by \rechargeAck. Daniele Di Pompeo and Michele Tucci were supported by \SoBigDataITAck. Luca Traini was supported by \icscAck. Vittorio Cortellessa was supported by \mattersAck.
\vspace{4em}

\bibliographystyle{elsarticle-num} 
\bibliography{bibliography}

\vspace{1em}
\noindent
\begin{wrapfigure}{l}{0.33\columnwidth}
  \centering
  \includegraphics[width=0.33\columnwidth]{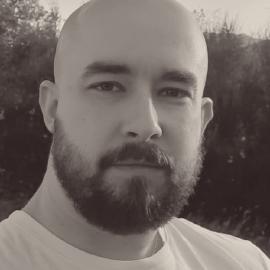}
\end{wrapfigure}
\textbf{Martin Beseda} is a Postdoctoral Researcher at the University of L'Aquila (UNIVAQ), Italy, where he focuses on detecting steady states in relation to algorithm runtime and advancing the use of quantum computers in computer science, chemistry, and physics. He earned Ph.D. in Computational Sciences and Plasma Engineering from the Technical University of Ostrava and Université Toulouse III - Paul Sabatier in 2022. Before UNIVAQ, he worked at LCPQ in Toulouse, ICGM in Montpellier, and IT4Innovations in Ostrava. His research has been published in prestigious journals, including Physical Review A, Plasma Sources Science and Technology, and Computational and Theoretical Chemistry.

\vspace{0.5em}
\noindent
\begin{wrapfigure}{l}{0.33\columnwidth}
  \centering
  \includegraphics[width=0.33\columnwidth]{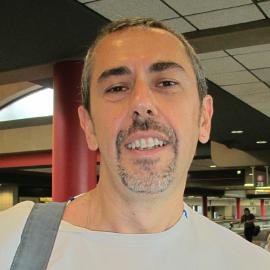}
\end{wrapfigure}
\textbf{Vittorio Cortellessa} is Full Professor at Department of Information Engineering, Computer Science, and Mathematics of University of L’Aquila (Italy). His research interests are in software performance and reliability, where he has experimented methodologies that span from model-driven engineering through multi-objective optimization to AI-based approaches. He has published about 150 papers on international conferences and journals in his areas of interest. He has received two 10-years most influential paper awards and five best paper awards. He has served in two editorial boards of journals, and he serves as chair and program committee member of leading conferences in his area of interest.

\vspace{0.5em}
\noindent
\begin{wrapfigure}{l}{0.33\columnwidth}
  \centering
  \includegraphics[width=0.33\columnwidth]{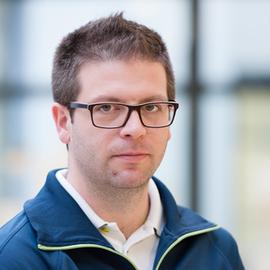}
\end{wrapfigure}
\textbf{Daniele Di Pompeo} is an Assistant Professor in the Department of Information Engineering, Computer Science, and Mathematics at the University of L’Aquila in Italy. He earned his Ph.D. in Information and Communication Technology (ICT) in 2019. His research focuses on model-based performance analysis, software refactoring, and search-based software engineering. He has published numerous articles in top-tier international journals and conferences related to his research. 

\vspace{0.5em}
\noindent
\begin{wrapfigure}{l}{0.33\columnwidth}
  \centering
  \includegraphics[width=0.33\columnwidth]{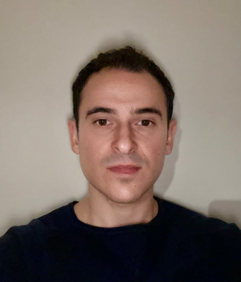}
\end{wrapfigure}
\textbf{Luca Traini} received his Ph.D. in Computer Science from the University of L’Aquila, Italy, in 2021, where he is currently an Assistant Professor. His research interests include performance engineering, software engineering and AI, and empirical software engineering. He has published papers in top-tier venues such as the IEEE/ACM International Conference on Automated Software Engineering (ASE), IEEE Transactions on Software Engineering (TSE), ACM Transactions on Software Engineering and Methodology (TOSEM), and Empirical Software Engineering (EMSE). In 2024, his work received the Industrial Paper Award at the IEEE International Conference on Software Analysis, Evolution and Reengineering (SANER).

\vspace{0.5em}
\noindent
\begin{wrapfigure}{l}{0.33\columnwidth}
  \centering
  \includegraphics[width=0.33\columnwidth]{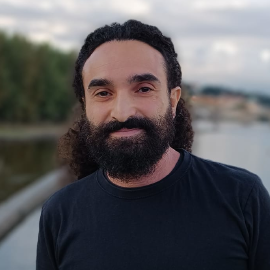}
\end{wrapfigure}
\textbf{Michele Tucci} is an Assistant Professor in the Department of Information Engineering, Computer Science and Mathematics at the University of L’Aquila, Italy. He earned his Ph.D. in Computer Science from the same university in 2021, under the supervision of Romina Eramo and Vittorio Cortellessa. From 2021 to 2023, he was a postdoctoral researcher in the Department of Distributed and Dependable Systems at Charles University, Prague. In 2023, he received the SPEC Impact Award for technical leadership. His research focuses on performance regression testing, software refactoring, and the optimization of software architectures w.r.t. quality attributes such as performance and reliability. 
 
\end{document}